\documentclass[11pt,twoside]{article}

\usepackage{asp2006}
\usepackage{epsf}
\usepackage{epsfig}
\usepackage{lscape}

\begin{document}
\title{The properties of unobscured AGN observed by XMM-Newton}   
\author{Stefano Bianchi,\altaffilmark{1,2} Matteo Guainazzi,\altaffilmark{2} Giorgio Matt,\altaffilmark{1} Nuria Fonseca Bonilla\altaffilmark{2}}   
\altaffiltext{1}{Dip. di Fisica, Universit\`a degli Studi Roma Tre, Via Vasca Navale 84, I-00146 Roma, Italy}
\altaffiltext{2}{XMM-Newton Science Operations Center, ESAC, ESA, Aptdo 50727, E-28080 Madrid, Spain}

\begin{abstract} 
Multiwavelength analysis on large samples of AGN provides an excellent tool to understand the physics of these objects. We present the largest catalog of XMM-\textit{Newton} targeted AGN, all with high SNR X-ray spectra. It includes all the radio-quiet objects observed by XMM-Newton, in targeted observations of the AGN panel. The principal X-ray properties of the catalog are complemented by multiwavelength data found in the literature (optical magnitudes, radio fluxes, H$\beta$ FWHM, BH masses). We present here some results on the correlation of these quantities. In particular, we find convincing evidence for an `Iwasawa effect' on the narrow component of the Fe K$\alpha$ line and a correlation between the soft-to-hard X-ray luminosity ratio and the H$\beta$ FWHM.
\end{abstract}



Our catalog consists of all the radio-quiet Type 1 AGN observed by XMM-\textit{Newton}, in targeted observations of the AGN panel, whose data are public as of February 2006. The total catalog comprises 130 radio quiet AGN. We refer the reader to Bianchi et al. (in prep.) for details on the data reduction and analysis. We present here two of the most interesting results of our study.

\begin{figure}
\begin{center}
\epsfig{file=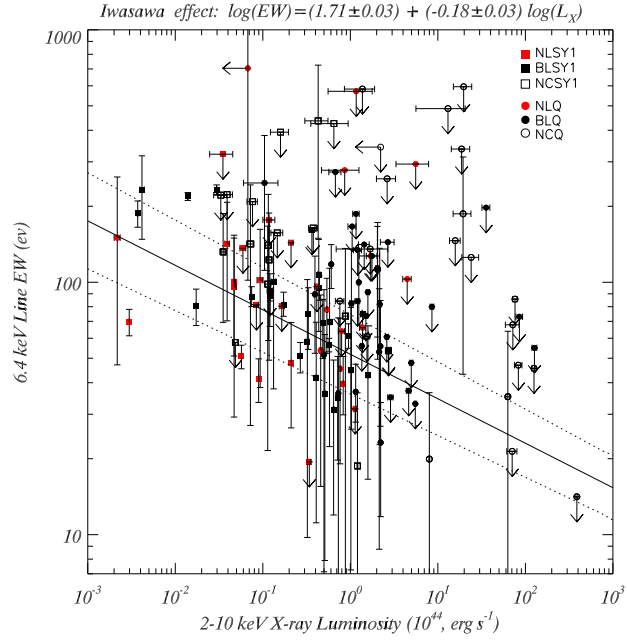, width=8.5cm}
\end{center}
\caption{\label{iwasawa}The `Iwasawa effect': Fe K$\alpha$ EW vs. X-ray luminosity. Objects are classified on the basis of their $M_{abs}$ and of the H$\beta$ FWHM, when available.}
\end{figure}

An anti-correlation between the equivalent width (EW) of the neutral Fe K$\alpha$ emission line and the 2-10 keV luminosity was found for the first time by Iwasawa \& Taniguchi (1993), according to \textit{Ginga} observations of 37 AGN. However, this result was later questioned by Jim\'enez Bail\'on et al. (2005) and Jiang et al. (2006), who underlined the importance of excluding radio-loud objects. We confirm here the `Iwasawa effect' (also known as `X-ray Baldwin effect') on a large number of genuine radio-quiet objects: $\log(EW)=(1.71\pm0.03)+(-0.18\pm0.03)\log(L_x)$ (see Fig. \ref{iwasawa}).

\begin{figure}
\begin{center}
\epsfig{file=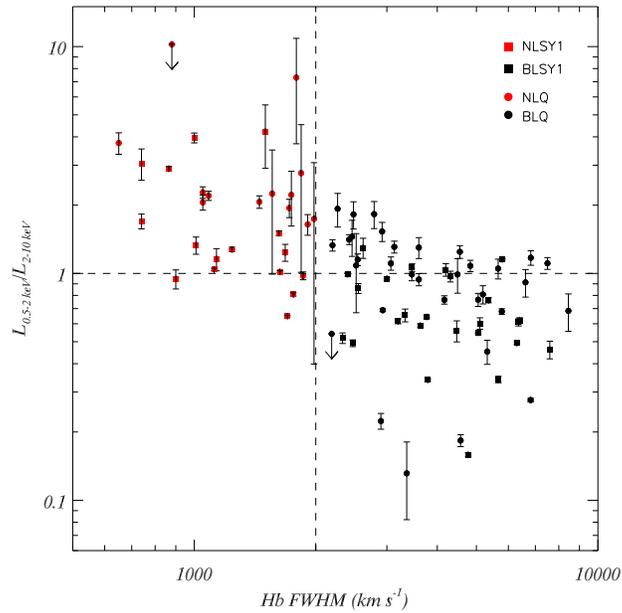, width=8.5cm}
\caption{\label{xratio}X-ray luminosity ratio vs H$\beta$ FWHM. The customary separation between narrow and broad line objects at $2\,000$ km s$^{-1}$ is plotted with a broken line, along with the limit $L_{0.5-2\,keV}$/$L_{2-10\,keV}>1$.}
\end{center}
\end{figure}

In 1996, Boller et al. showed that the soft X-ray photon index of AGN correlates with the FWHM of their optical lines. In particular, steeper indexes where found for narrower H$\beta$ emission lines. In Fig. \ref{xratio}, we show a less model-dependent parameter, which is the ratio between the 0.5-2 keV luminosity and the 2-10 keV luminosity: sources with narrower H$\beta$ FWHM tends to have larger $L_{0.5-2\,keV}$/$L_{2-10\,keV}$.

\end{document}